\documentclass[aps,prl,superscriptaddress,longbibliography]{revtex4-2}

\usepackage{gensymb}
\usepackage{color}
\usepackage[normalem]{ulem}
\usepackage{graphicx}
\usepackage{dcolumn}
\usepackage{float}
\usepackage{bm}
\usepackage{amsmath}
\usepackage{comment}

\newcommand{\be}{\begin{equation}}
\newcommand{\ee}{\end{equation}}
\newcommand{\bfig}{\begin{figure}}
\newcommand{\efig}{\end{figure}}

\newcommand{\NRA}{Nd$_3$Ru$_4$Al$_{12}$}
\newcommand{\GRA}{Gd$_3$Ru$_4$Al$_{12}$}
\newcommand{\DRA}{Dy$_3$Ru$_4$Al$_{12}$}
\newcommand{\TRA}{Tb$_3$Ru$_4$Al$_{12}$}
\newcommand{\RRA}{$R_3$Ru$_4$Al$_{12}$}
\newcommand{\URA}{U$_3$Ru$_4$Al$_{12}$}
\newcommand{\PRA}{Pr$_3$Ru$_4$Al$_{12}$}

\newcommand{\RX}{\textit{R}$_8$Co\textit{X}$\kern-0.15em_{3}$}

\begin{document}
\title{Collinear ferromagnetism with reduced moment length in kagome magnet \NRA{}}

\author{Yuki Ishihara}
\affiliation{Department of Applied Physics and Quantum-Phase Electronics Center (QPEC), The University of Tokyo, Bunkyo, Tokyo 113-8656, Japan}

\author{Ryota Nakano}
\affiliation{Department of Applied Physics and Quantum-Phase Electronics Center (QPEC), The University of Tokyo, Bunkyo, Tokyo 113-8656, Japan}

\author{Rinsuke Yamada}
\affiliation{Department of Applied Physics and Quantum-Phase Electronics Center (QPEC), The University of Tokyo, Bunkyo, Tokyo 113-8656, Japan}

\author{Takuya Nomoto}
\affiliation{Department of Physics, Tokyo Metropolitan University, Hachioji, Tokyo 192-0397, Japan}

\author{Priya R. Baral}
\affiliation{Department of Applied Physics and Quantum-Phase Electronics Center (QPEC), The University of Tokyo, Bunkyo, Tokyo 113-8656, Japan}

\author{Moritz M. Hirschmann}
\affiliation{RIKEN Center for Emergent Matter Science (CEMS), Wako, Saitama 351-0198, Japan}

\author{Kamini Gautam}
\affiliation{RIKEN Center for Emergent Matter Science (CEMS), Wako, Saitama 351-0198, Japan}

\author{Kamil K. Kolincio}
\affiliation{RIKEN Center for Emergent Matter Science (CEMS), Wako, Saitama 351-0198, Japan}
\affiliation{Faculty of Applied Physics and Mathematics, Gda{\'n}sk University of Technology, Narutowicza 11/12, 80-233 Gda{\'n}sk, Poland}

\author{Akiko Kikkawa}
\affiliation{RIKEN Center for Emergent Matter Science (CEMS), Wako, Saitama 351-0198, Japan}

\author{Seno Aji}
\affiliation{The Institute for Solid State Physics, The University of Tokyo, Kashiwa 277-8581, Japan}
\affiliation{Present address: Department of Physics, Faculty of Mathematics and Natural Sciences, Universitas Indonesia, Depok, 16424, Indonesia}

\author{Hiraku Saitoh}
\affiliation{The Institute for Solid State Physics, The University of Tokyo, Kashiwa 277-8581, Japan}

\author{Masaaki Matsuda}
\affiliation{Neutron Scattering Division, Oak Ridge National Laboratory, Oak Ridge, TN, 37831, USA}


\author{Yasujiro Taguchi}
\affiliation{RIKEN Center for Emergent Matter Science (CEMS), Wako, Saitama 351-0198, Japan}

\author{Taka-hisa Arima}
\affiliation{Department of Advanced Materials Science, The University of Tokyo, Kashiwa 277-8561, Japan}
\affiliation{RIKEN Center for Emergent Matter Science (CEMS), Wako, Saitama 351-0198, Japan}

\author{Yoshinori Tokura}
\affiliation{RIKEN Center for Emergent Matter Science (CEMS), Wako, Saitama 351-0198, Japan}
\affiliation{Department of Applied Physics and Quantum-Phase Electronics Center (QPEC), The University of Tokyo, Bunkyo, Tokyo 113-8656, Japan}
\affiliation{Tokyo College, The University of Tokyo, Tokyo 113-8656, Japan.}

\author{Taro Nakajima}
\affiliation{The Institute for Solid State Physics, The University of Tokyo, Kashiwa 277-8581, Japan}
\affiliation{Institute of Materials Structure Science, High Energy Accelerator Research Organization, Tsukuba 305-0801, Japan}
\affiliation{RIKEN Center for Emergent Matter Science (CEMS), Wako, Saitama 351-0198, Japan}

\author{Max Hirschberger}
\thanks{hirschberger@ap.t.u-tokyo.ac.jp}
\affiliation{Department of Applied Physics and Quantum-Phase Electronics Center (QPEC), The University of Tokyo, Bunkyo, Tokyo 113-8656, Japan}
\affiliation{RIKEN Center for Emergent Matter Science (CEMS), Wako, Saitama 351-0198, Japan}

\date{\today}

\begin{abstract}
We determine the magnetic ground state of the kagome lattice magnet \NRA{} by single-crystal neutron diffraction, supported by experiments with polarized neutrons. We identify this material as a collinear ferromagnet ('hex-FM') with uniform moment length $m_c\approx 2.1\,\mu_\mathrm{B}$/Nd and ordering vector $\bm{Q} = 0$,
in contrast to a previous, seminal report that proposed unequal moment lengths on two Nd sites -- here called the 'ortho-FM' state. Our analysis of the flipping ratio in polarized neutron scattering is consistent with the hex-FM state. The results provide a microscopic basis for understanding the large fluctuation-induced Hall and Nernst responses near $T_\mathrm{C} \approx 41\,$K, as previously reported for \NRA{}.
\end{abstract}

\maketitle

\section{Introduction}

Materials with a kagome lattice of magnetic ions have served as testing grounds for theories of quantum spin liquid phases~\cite{Katsura2010, Norman2016}, topological spin waves~\cite{Chisnell2015,Hirschberger2015,Lee2015} and electronic flat band or Dirac states~\cite{Ye2018,Checkelsky2024}. Complex magnetic ordering has also been predicted for the kagome lattice, induced by frustrated or competing interactions and further stabilized by spin-orbit interactions~\cite{Ozawa2022, Watanabe2022, Lin2024}. In fact, even a simple, collinear ferromagnet can acquire a net spin-winding or scalar spin chirality under the influence of thermal fluctuations~\cite{Hou2017, Wang2019, Carnahan2021}. The kagome lattice geometry promotes this fluctuation effect~\cite{Kolincio2021, Kolincio2023}, driving strong electron- or magnon-driven responses due to non-cancellation of spin chirality contributions from various lattice plaquettes~\cite{Ideue2012}. Some of us have previously studied the electronic properties of \NRA{}, where the inversion center is located between two kagome lattice layers and where local Dzyaloshinskii-Moriya interactions are allowed [Fig.~\ref{Fig1}(a)]. In this metal, experiments show large electronic and thermoelectric responses that are attributed to thermal spin fluctuations with a net chirality~\cite{Kolincio2021, Kolincio2023}. The fluctuation mechanism is illustrated in Fig.~\ref{Fig1}(c).

Based on high-quality neutron diffraction data on single crystals, previous studies argued that the magnetic ground state in \NRA{} and \PRA{} breaks both six- and three-fold rotation symmetries, realizing a collinear ferromagnetic state with two different Wyckoff sites for the Nd (or Pr) ion~\cite{Gorbunov2016, Henriques2018_Pr}. The magnetic space group of \NRA{} reported in Ref.~\cite{Gorbunov2016} is orthorhombic, thus we call this configuration the ‘ortho-FM’ state. The magnetic moments on the two rare earth sites in ortho-FM were reported to differ significantly, $2.66\,\mu_\mathrm{B}$ and $0.92\,\mu_\mathrm{B}$, respectively. However, ultrasound attenuation studies did not reveal any evidence for the existence of two distinct Nd sites~\cite{Ishii2018,Suzuki2020} and detailed crystal field calculations could not reproduce the observed value of the net magnetization~\cite{Ishii2020}. We note that bulk magnetization measurements show a net magnetization of $\sim2.1\,\mu_\mathrm{B}$/Nd along the $c$-axis~\cite{Gorbunov2016,Kolincio2021}, which is not consistent with the free-ion moment for Nd, $3.6\,\mu_\mathrm{B}$/Nd, nor with the fluctuating moment observed in the Curie-Weiss behavior of \NRA{} at high $T$~\cite{Ge2014, Gorbunov2016, Ishii2020}. 

These seemingly inconsistent experiments, and the strong response in electric and thermoelectric transport, motivated us to re-examine the magnetic ground state of \NRA{} by neutron diffraction. We quantitatively compare integrated magnetic intensity to various magnetic structure models and also conduct a careful analysis of the flipping ratio in polarized-neutron scattering. Thus, we demonstrate that \NRA{} is a simple collinear ferromagnet (hex-FM) with no detectable in-plane canting and equal moment length on all Nd sites. 

Our results highlight the sharp difference between the lighter (Pr, Nd) and heavier (Gd, Tb, Ho and so on) rare earths in the \RRA{} series. The former have a significant net magnetization and dominantly collinear magnetic order~\cite{Gladyshevskii1993,Chandragiri2016}. These are outliers in Fig.~\ref{Fig1}(b), while the heavier $R$ follow de-Gennes scaling $T_\mathrm{N}\sim(g_J-1)^2J(J+1)$ consistent with a Ruderman-Kittel-Kasuya-Yosida (RRKY) picture for the magnetic interactions; here, $J$ is the total angular momentum of the $4f$ shell and $g_J$ is the Land{\'e} $g$-factor~\cite{JensenMackintosh_book,deGennes_RKKY}. 


For context, we briefly review the complex magnetism of the heavier \RRA{} in Fig.~\ref{Fig1}(b), as studied by X-ray and neutron diffraction. \GRA{} hosts spin spirals~\cite{Nakamura2018, Matsumura2019,Nakamura2023} and magnetic skyrmions~\cite{Hirschberger2019, Hirschberger2021,Hirschberger2024}; by band filling control, the spin spiral can be commensurably locked to the underlying atomic lattice, thus satisfying the symmetry constraints for a $p$-wave magnet with parity-odd spin splitting of the Fermi surface~\cite{Hellenes2023, Chakraborty2025, Yamada2025}. In \DRA{}, neutron diffraction demonstrates a complex spin texture composed of magnetic trimers with all-in-all-out or vortex-like patterns~\cite{Gorbunov2014,Gao2019}. The magnetic order of $R = \,$Ho~\cite{Gorbunov2018} and $R=\,$Yb~\cite{Nakamura2015,Sato2018} remains underexplored at present. Finally, in the $5f$ system \URA{} with the same crystal structure, a noncollinear, XY-type antiferromagnetic state has been reported~\cite{Troc2012,Asaba2020}. 


\section{Sample preparation}

Single crystals were grown by the Czochralski method in an Ar atmosphere ($p\approx 1\,$atm) with slight excess of Nd and Al to compensate for evaporation and oxidation during the growth. The samples were characterized by powder X-ray diffraction and single-crystal Laue diffraction. Through inspection under a microscope with nicol prism, pieces with an Al-rich impurity phase were excluded. The magnetization and electronic transport response of these samples is reported in Ref.~\cite{Kolincio2021}. 

\section{Unpolarized neutron diffraction}

\subsection{Experimental setup}
We use beamline PONTA-5G at the Japan Research Reactor 3 (JRR-3) for neutron diffraction on an aligned and polished single crystal of mass $m \approx 100\,$mg (Sample 1). 

The beam geometry for unpolarized scattering in two-axis mode is shown in Fig.~2(a) inset, with incoming and outgoing neutron beams ($\bm{k}_i$, $\bm{k}_f$) in the $(HK0)$ scattering plane. We use thermal neutrons with $E = 34.05\,$meV for this experiment, with the $(002)$ reflection of graphite as a monochromatizer. The order parameter is $\bm{Q}=0$~\cite{Gorbunov2016} so that magnetic and nuclear intensities coincide at the same positions of reciprocal space. To subtract the nuclear intensity, we record Bragg intensities at $T=2\,$K and at $T=120\,$K and compute $I_{\mathrm{mag}}(\bm{Q}) = I_{2\,\mathrm{K}} - I_{120\,\mathrm{K}}$. The results of this procedure and respective error bars are shown in Fig.~\ref{Fig2}(a,b). We propagate statistical errors from the fitting of Gaussian-shaped intensity in $\omega-2\theta$ scans and impose a conservative lower bound of $5\,\%$ of $I_{2\,\mathrm{K}}$ for the uncertainty. Here, $2\theta$ and $\omega$ are the scattering angle and the (rocking) angle of the sample surface relative to the incoming beam, respectively.

\subsection{Refinement method for neutron scattering intensities}
The nuclear and magnetic structure factors are calculated, respectively, from the crystallographic structure of the sample and from its magnetic (dipole moment) pattern:
\begin{align}
F^\mathrm{nuc}_\mathrm{calc}(\bm{Q}) &= \sum_{j\in  \mathrm{u.c.}}b_j e^{\imath \bm{Q}\cdot\bm{r}_j}\\
\label{eq:magnetic_structure_factor}
\bm{F}^\mathrm{mag}_\mathrm{calc}(\bm{Q}) & = -\left(\frac{\gamma r_0}{2}\right) \sum_{j\in  \mathrm{m.u.c.}} \bm{m}_j\,f_j^\mathrm{mag}(\left|\bm{Q}\right|)\, e^{\imath \bm{Q}\cdot\bm{r}_j}
\end{align}
and $b_j$, $\bm{r}_j$, $f_j^\mathrm{mag}(\left|\bm{Q}\right|)$, and $\bm{m}_j$ are the nuclear scattering length, the position of an atomic nucleus, the magnetic form factor, and the magnetic moment at site $j$, respectively; m.u.c (u.c.) refers to the magnetic unit cell (crystallographic unit cell). Moreover, $\gamma$ and $r_0$ are the gyromagnetic ratio and the classical electron radius, respectively. We include a correction for the isotropic extinction effect, assume $100\,\%$ occupation of all atomic sites, set the (isotropic) atomic displacement factors to zero, and neglect absorption of neutrons in the sample~\cite{Squires2012}.

For the magnetic structure factor, we consider only contributions from rare earth Nd ions, normalize the magnetic moment $\bm{m}_j$ to unit length and use $\bm{F}_\mathrm{calc}^{\mathrm{mag},\perp}=\bm{F}_\mathrm{calc}^\mathrm{mag}-(\hat{\bm{Q}}\cdot \bm{F}_\mathrm{calc}^\mathrm{mag})\hat{\bm{Q}}$ for $\hat{\bm{Q}} = \bm{Q}/ \left|\bm{Q}\right|$. Using a custom software~\cite{Akatsuka2024}, we calculate the nuclear and magnetic structure factors as $I^\mathrm{nuc}_\mathrm{calc} =\Phi \left|F_\mathrm{calc}^\mathrm{nuc}\right|^2$ and $I^\mathrm{mag}_\mathrm{calc} =\Phi \left|\bm{F}_\mathrm{calc}^{\mathrm{mag},\perp}/x\right|^2$ where $\Phi$ and $x$ are the neutron flux and the ratio $V_\mathrm{m.u.c}/V_\mathrm{u.c.}$ of the volume of magnetic and nuclear unit cells. In the case of $N$ (equally populated) domains, the intensity is calculated for each domain $k$ and an average is obtained as $I^\mathrm{mag}_\mathrm{calc}= (1/N)\sum_{k=1}^N I^\mathrm{mag}_{\mathrm{calc},k}$.

The observed structure factor is $F_\mathrm{obs} = \sqrt{I\sin2\theta}$ with the scattering angle $2\theta$ and scattering intensity $I$. We obtain two scale factors $F_\mathrm{obs}^\mathrm{nuc}=s_\mathrm{N} \left|F^\mathrm{nuc}_\mathrm{calc}\right|$ and $F_\mathrm{obs}^\mathrm{mag}=s_\mathrm{M} \left|\bm{F}^\mathrm{mag}_\mathrm{calc}\right|$ by minimizing the 
weighted $\chi^2$ value.
The magnetic moment length is calculated as $m=s_\mathrm{M}/s_\mathrm{N}$. In the nuclear refinement, we replace 
\begin{equation}
F_\mathrm{calc} \rightarrow F_\mathrm{calc} / \left(1+ \eta_\mathrm{ext}\left|F_\mathrm{calc}\right|^2/\sin2\theta\right)^{{1/4}}
\end{equation}
where $\eta_\mathrm{ext}$ is the (isotropic) extinction parameter. Then, $\eta_\mathrm{ext}$ is held fixed in the magnetic refinement. For the analysis of magnetic scattering, we use only those reflections where the error of $I_\mathrm{obs}^\mathrm{mag}$ is smaller than two times $I_\mathrm{obs}^\mathrm{mag}$.
\subsection{Refinement of intensities in ferromagnetic structure}

Figure~\ref{Fig2} shows large nuclear intensity (grey) at $(-240)$ and $(-480)$, but relatively large magnetic intensity (blue) at $(-220)$, $(-230)$, for example. From this neutron experiment, we determine the lattice constants of \NRA{} as $a = 8.887(5)\,$\AA{} and $c=9.647(5)\,$\AA{} at base temperature. In Fig.~\ref{Fig3}(a), we compare the observed and calculated structure factors for the collinear ferromagnetic model. We find good agreement in this $(HK0)$ scattering plane with no free parameters (besides $s_\mathrm{M}$); additional single crystal neutron diffraction in the $(H0L)$ and $(HHL)$ planes, on  different samples, further confirms our conclusion (Ref.~\cite{SI}). Figure~\ref{Fig3}(b) compares calculated and observed structure factors for the nuclear scattering intensity, where the intracell coordinates of Nd, Ru, Al (as well as the isotropic extinction parameter) are the only important adjustable parameters. The results of the crystal structure refinement are summarized in the inset text of Fig.~\ref{Fig3}(b). The ordered magnetic moment is $m_c=2.47\,\mu_\mathrm{B}/$Nd, consistent with bulk magnetization experiments~\cite{Gorbunov2014}.

\subsection{Consideration about $(100)$ reflection}

Figure~\ref{Fig4}(d) shows that the magnetic intensity at $(100)$ is much weaker than at $(200)$. This is well consistent with the proposed hex-FM state, but -- as we show in the following -- immediately rules out various possible magnetic orders of lower symmetry, including ortho-FM. 

The hex-FM state in Fig.~\ref{Fig4}(a) has magnetic space group $P6_3/m m^\prime c^\prime$ and $24$ individual symmetry operations. In Appendix~\ref{appendix_section_mSGs}, we consider six magnetic subspacegroups of $P6_3/mm^\prime c^\prime$ with $\bm{Q} = 0$, with magnetic moment on the $6h$ Wyckoff site: one of them is a ferromagnetic state with different magnetization on the two kagome layers in Fig.~\ref{Fig1}(a). Four are canted ferromagnets, where the in-plane component of the magnetic moment has "in/out" or "vortex-like" patterns on individual triangles of the kagome lattice. These five have $12$ symmetries each. Finally, there is an orthorhombic subspacegroup with $8$ symmetries, the ortho-FM state with two Nd-sites depicted in Fig.~\ref{Fig4}(c). Among these six, only the "in/out" patterns are consistent with weak $(100)$ intensity [Fig.~\ref{Fig4}(b)]; for them, the sum of in-plane moments in a plane perpendicular to $\bm{Q}_{(100)}$ is fully parallel to $\bm{Q}_{(100)}$ (thus, not detectable). However, even the "in/out" canting is further ruled out in Appendix~\ref{appendix_section_mSGs}.

Considering the good agreement of experiment and the hex-FM model, it is not necessary to further lower the magnetic space group symmetry to describe our data. To bring out this point, we found it helpful to compare ortho-FM in magnetic space group (mSG) $Cm^\prime c^\prime m$ to hex-FM [Fig.~\ref{Fig4}(a,b)] in mSG $P6_3/m^\prime m^\prime c$. 
Figure~\ref{Fig4}(d) contrasts the observed magnetic intensities along $(H00)$ for the two models.
As noted above, hex-FM gives a better description of the $(100)$ reflection, which is rather sensitive to different spin configuration on the "upper" and "lower" kagome layer in Fig.~\ref{Fig1}(a). At $(100)$, the experimentally observed scattering intensity is about four times smaller than predicted for ortho-FM. We use the moment values $m_{\mathrm{Nd-1}}=2.66\,\mu_\mathrm{B}$ and $m_{\mathrm{Nd-2}}=0.95\,\mu_\mathrm{B}$ for two inequivalent Nd sites in orho-FM, as reported in Ref.~\cite{Gorbunov2016}. We also carefully account for six magnetic domains (three pairs of time-reversal partnered domains) in ortho-FM.

In Fig.~4(d), there is finite magnetic intensity at $(100)$, contrary to the prediction of the hex-FM model. We assign this slight excess of intensity to the $\lambda/2$ effect, as $I_\mathrm{obs}^\mathrm{mag}(100)/I_\mathrm{obs}^\mathrm{mag}(200)$ is on the order of $3\sim4\,\%$.

\section{Diffraction experiments with polarized neutrons}

\subsection{Experimental Method}
For polarized neutron diffraction, we set beamline PONTA-5G in triple axis mode and use the $(111)$ reflection of a Cu$_2$MnAl Heusler ferromagnet as both monochromatizer and analyzer plate. The sample is again mounted in the $(HK0)$ plane. Using a spin flipper, the spin of the incoming neutrons can be switched between "up" and "down" configurations -- parallel and antiparallel to the $c$-axis of \NRA{}. This experimental condition is depicted in Fig.~\ref{Fig5}(a). A magnetic field of $B = 1\,$T is applied at the sample position to suppress ferromagnetic domain patterns and to avoid inhomogeneous stray fields, which can artificially depolarize the neutron beam. Corrections for half-$\lambda$ imperfections of the monochromator and for depolarization of the neutron beam ($1-p = 2.4\,\%\pm 1\%$) are discussed in Appendix~\ref{appendix:polarized_neutron}. 

\subsection{Excluding in-plane canting of magnetic moments by polarized neutron scattering}

Rocking ($\omega$) scans in the ``spin flipper ON/OFF'' configurations [Figs.~5(b,c)] allow us to track the experimentally observed flipping ratio
\begin{equation}
P_{\mathrm{obs}} = \frac{I_\mathrm{du}-I_\mathrm{uu}}{I_\mathrm{du}+I_\mathrm{uu}},
\end{equation}
which we compare to calculations of $P_{\mathrm{calc}}$ for hex-FM including the measured beam depolarization. We note that even a fully collinear state, with all moments along the $c$-axis, can induce $I_\mathrm{du}$ intensity due to beam depolarization $1-p$. Further, $I_\mathrm{uu}$ can be suppressed due to partial cancellation between nuclear and magnetic scattering, so that a flipping ratio $P<1$ is possible~(Appendix~\ref{appendix:polarized_neutron}).

In Fig.~\ref{Fig5}(d), the agreement between $P_\mathrm{obs}$ and $P_\mathrm{calc}$ is good for $m_c^\mathrm{opt} \approx 2.3\,\mu_\mathrm{B}$, and we observe no systematic deviations from $P_\mathrm{obs} = P_\mathrm{calc}$ that could be indicative of a canted ($ab$ plane) component of the magnetic moment. This means that any static canting of the magnetic moments (Appendix~\ref{appendix_spacegroups}) is below our detection threshold. Although it is true that the refined $m_c^\mathrm{opt}$ exceeds the moment value from bulk magnetization measurements by about $10\,\%$~\cite{Gorbunov2016}, this is reasonable considering possible systematic variations of $p$.

\section{Discussion}

As Ref.~\cite{Gorbunov2016} also emphasizes, the ordered magnetic moment of $R=\,$Nd is substantially reduced ($\sim30\,$\%) from the free-ion value. For further insight into the magnetism of \NRA{}, we performed electronic structure calculations using the Vienna Ab initio Simulation Package (VASP)~\cite{Kresse1993,Kresse1999}. Here, the ortho-FM state is generally unstable and converges to hex-FM. We employed a $6 \times 6 \times 6$ Monkhorst-Pack $\bm{k}$-grid, a plane-wave cutoff energy of $400\,$eV, the GGA exchange–correlation functional of Perdew, Burke, and Ernzerhof~\cite{Perdew1996}, and included spin-orbit coupling. Hund’s third rule for early rare-earth elements stipulates that $L$ and $S$ prefer to align antiparallel and that the orbital (spin) angular momentum in the $4f$ shell of Nd is $L = 3$ ($S=3/2$), corresponding to $m_L = +6\,\mu_\mathrm{B}$ ($m_S = -3\,\mu_\mathrm{B}$). The electronic structure calculation with Hubbard $U = 0$ and Hund's coupling $J_\mathrm{H}=0$ confirms the suppression of the orbital magnetic moment in the hex-FM state, yielding $m_S = 3.0\,\mu_\mathrm{B}$ and $m_L =-1.4\,\mu_\mathrm{B}$. An increase in $U$ or $J_\mathrm{H}$ changes $m_S$ but little, while $m_L$ moves to larger negative values.

Supporting this discussion, our polarized and unpolarized neutron scattering data are fully consistent with the hex-FM model for \NRA{}, without any need for further symmetry lowering.
This finding resolves an apparent tension between an earlier neutron diffraction study that proposed the ortho-FM state and subsequent research on phononic properties and crystal field effects, which did not report any evidence for inequivalent Nd sites. 
Contrary to antiferromagnetic $R =\,$Gd, Tb, Ho and so on, \RRA{} with $R=\,$Nd is thus a collinear ferromagnetic metal, yet still hosts strong transport responses induced by thermal fluctuations. While the spins are static and aligned along the $c$-axis at base temperature, thermal fluctuations under the influence of Dzyaloshinskii-Moriya interactions on individual spin-trimers induce a net scalar spin chirality, which strongly impacts the flow of conduction electrons in \NRA{} at and above $T_\mathrm{C}$~\cite{Kolincio2021,Kolincio2023}.


\section{Conclusions}

Single-crystal neutron diffraction, including polarized neutron scattering, demonstrates that \NRA{} is a collinear $\bm{Q}=0$ ferromagnet with uniform direction of the spin magnetic moment along the crystallographic $c$-axis. Comparing nuclear and magnetic scattering, we calculate $m_c \approx 1.9-2.3\,\mu_\mathrm{B}$ per Nd with no evidence for in-plane canting of the magnetic moments. 
Our analysis illustrates the stark difference between \NRA{} and antiferromagnets such as \GRA{}, \DRA{} and \TRA{} in this rare earth intermetallic family, consistent with the violation of de-Gennes scaling (Fig. 1). The data also support the notion that thermal fluctuations -- around a simple ferromagnetic ground state -- strongly affect electronic transport properties~\cite{Hou2017} of \NRA{}, when  $T\approx T_\mathrm{C}$: With a finite magnetic field $B$ applied along the $c$-axis, the spin expectation value $\left<\bm{S}_i\right>$ is collinear; nevertheless, the thermal and lattice average of the scalar spin chirality $\left<\bm{S}_i\cdot(\bm{S}_j\times\bm{S}_k)\right>$ for neighboring sites $i$, $j$, $k$ has a dramatic impact on the electrical Hall effect and thermoelectric Nernst effect, due to the kagome lattice geometry~\cite{Kolincio2021,Kolincio2023}.\\

\textbf{Acknowledgements}\\
We thank R.\ Misawa for helpful discussions. P.~R.~B.\ acknowledges Swiss National Science Foundation (SNSF) Postdoc.Mobility grant P500PT\_217697 for financial assistance. M.~M.~H.\ was funded by the Deutsche Forschungsgemeinschaft (DFG, German Research Foundation) - project number 518238332 and by the RIKEN Special Postdoctoral Researcher Program. M.~H.\ is supported by the Deutsche Forschungsgemeinschaft (DFG, German Research Foundation) via Transregio TRR 360 – 492547816. The neutron scattering experiment at beamline PONTA-5G of JRR-3 was carried out under the proposal No. 24506. A portion of this research used resources at the High Flux Isotope Reactor, a DOE Office of Science User Facility operated by the Oak Ridge National Laboratory. The beam time was allocated to HB-1 on proposal number IPTS-30717.1.
This work was supported by JSPS KAKENHI Grant Nos. JP23H05431, JP24H01607, JP22K20348, JP23K13057, JP24H01604, JP25K17336, JP26H00644 and JP26H01290. This work was partially supported by the Japan Science and Technology Agency via JST CREST Grant Number JPMJCR20T1, JST PRESTO JPMJPR259A, and JST FOREST JPMJFR2238. It was also supported by Japan Science and Technology Agency (JST) as part of Adopting Sustainable Partnerships for Innovative Research Ecosystem (ASPIRE), Grant Number JPMJAP2426.\\

\bibliography{Nd3Ru4Al12}

\clearpage
\newpage

\begin{figure}[tb]
  \begin{center}
		\includegraphics[clip, trim=0cm 0cm 0cm 0cm, width=0.8\linewidth]{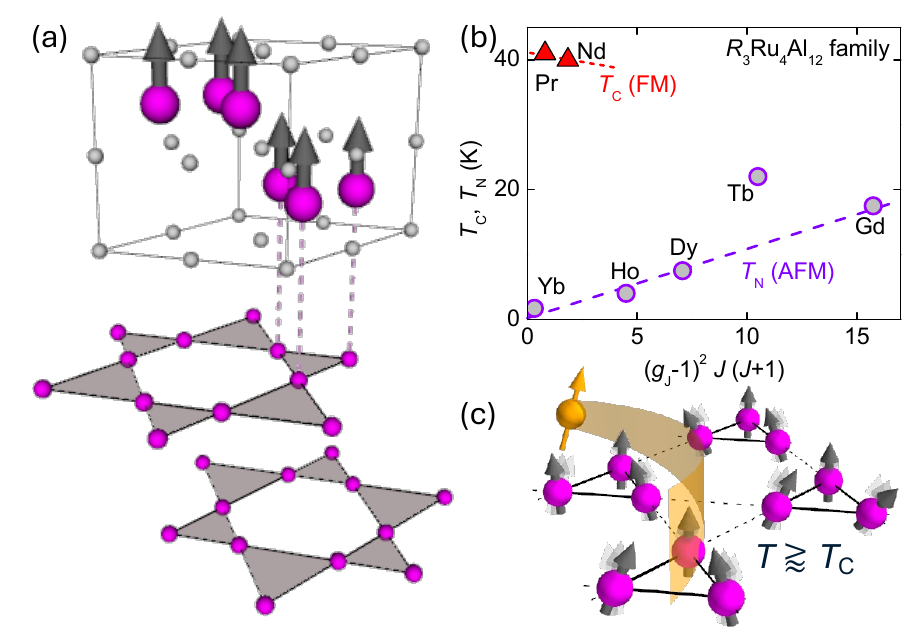}
    \caption[]{(color online). Ferromagnetic order in \RRA ($R =\,$Pr, Nd) with kagome lattice structure for the rare earth sublattice ('hex-FM'). (a) Ferromagnetic ordering as reported here for $R  =\,$Nd (hex-FM). The staggered kagome sublattice of Nd is depicted by the projection. Magenta (Nd) and grey (Ru) atoms, but not Al atoms, are shown. (b) Ferromagnetic ($T_\mathrm{C}$) and antiferromagnetic ($T_\mathrm{N}$) transition temperatures of \RRA{} as a function of the de-Gennes factor. For conduction-electron mediated magnetic interactions, we expect a linear relation. $R =\,$Pr, Nd are clear outliers with a sizable net magnetization. (c) Fluctuation-induced Hall effect in \NRA{} at and above $T_\mathrm{C}$, as reported in Ref.~\cite{Kolincio2021}. The moving conduction electron (yellow) is deflected by an emergent magnetic field due to time-averaged scalar spin chirality, although the thermal excitation value of each spin is collinear.
}
    \label{Fig1}
  \end{center}
\end{figure}

\begin{figure*}[tb]
  \begin{center}
    \includegraphics[clip, trim=0cm 0cm 0cm 0cm, width=0.85\linewidth]{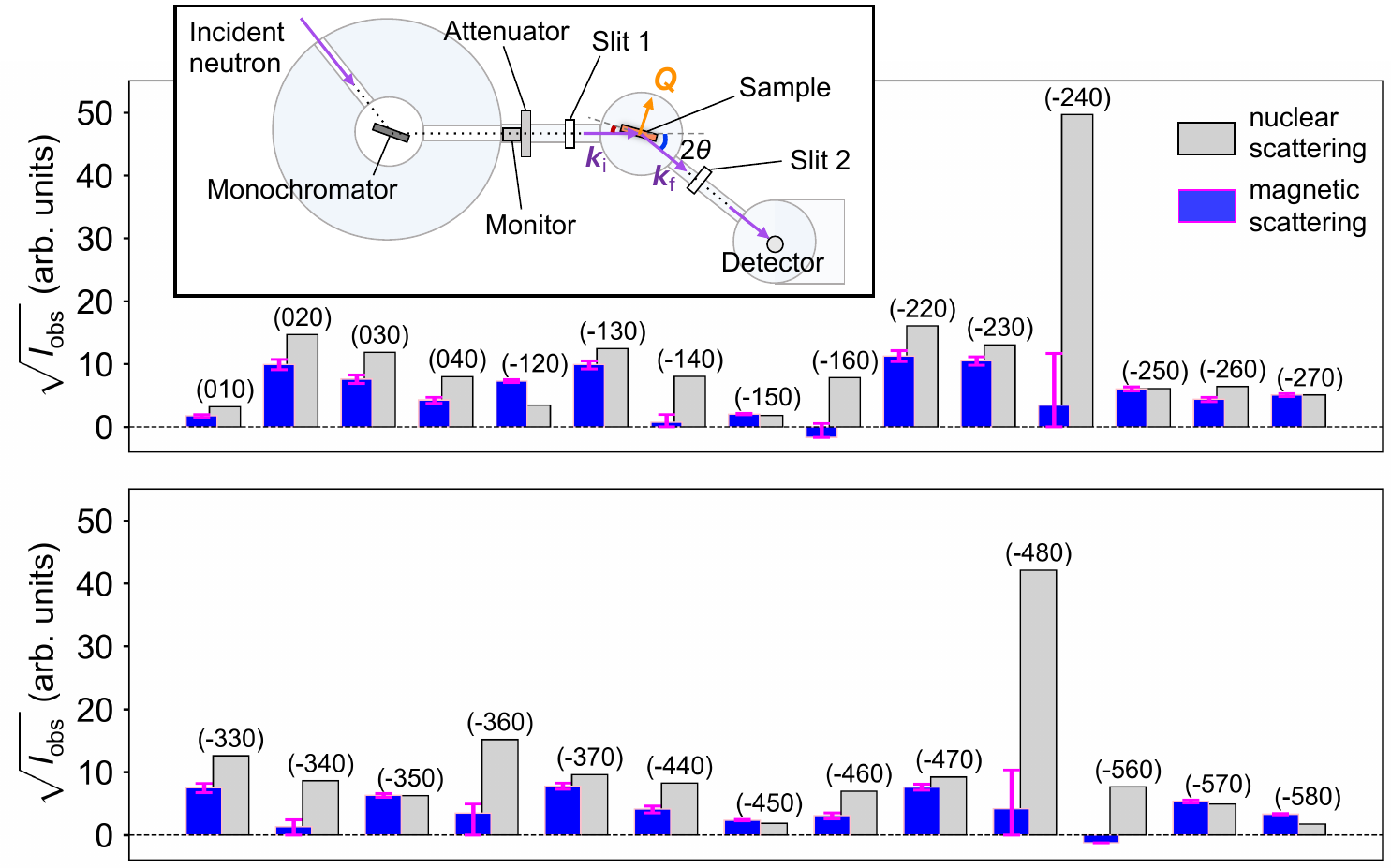}
    \caption[]{(color online). Unpolarized neutron scattering on \NRA{}. The nuclear scattering intensity $I_\mathrm{nuc}$ from the crystallographic structure (grey, $T = 120\,$K), magnetic scattering intensity $I_\mathrm{mag}$ (blue), and error of the magnetic intensity $\Delta I_\mathrm{mag}$ (magenta bars) are indicated. The Miller indices $(HKL)$ are indicated above each dataset. We obtain $I_\mathrm{mag}$ from $T = 2\,$K diffraction data $I_{2\,\mathrm{K}}$ by subtracting $I_\mathrm{nuc}$. The square root is taken for the absolute value of $I_\mathrm{obs}$ and multiplied by the sign of $I_\mathrm{obs}$. The error bars are calculated as a statistical uncertainty of the fit to $\omega-2\theta$ scans, but required to be larger than $5\,\%$ of $I_{2\,\mathrm{K}}$.
 (a), Inset: Geometry of beamline PONTA-5G of JRR-3 for this experiment in two-axis geometry. The $(HK0)$ crystal plane is horizontal. 
}
    \label{Fig2}
  \end{center}
\end{figure*}

\begin{figure}[tb]
  \begin{center}
		\includegraphics[clip, trim=0cm 0cm 0cm 0cm, width=0.6\linewidth]{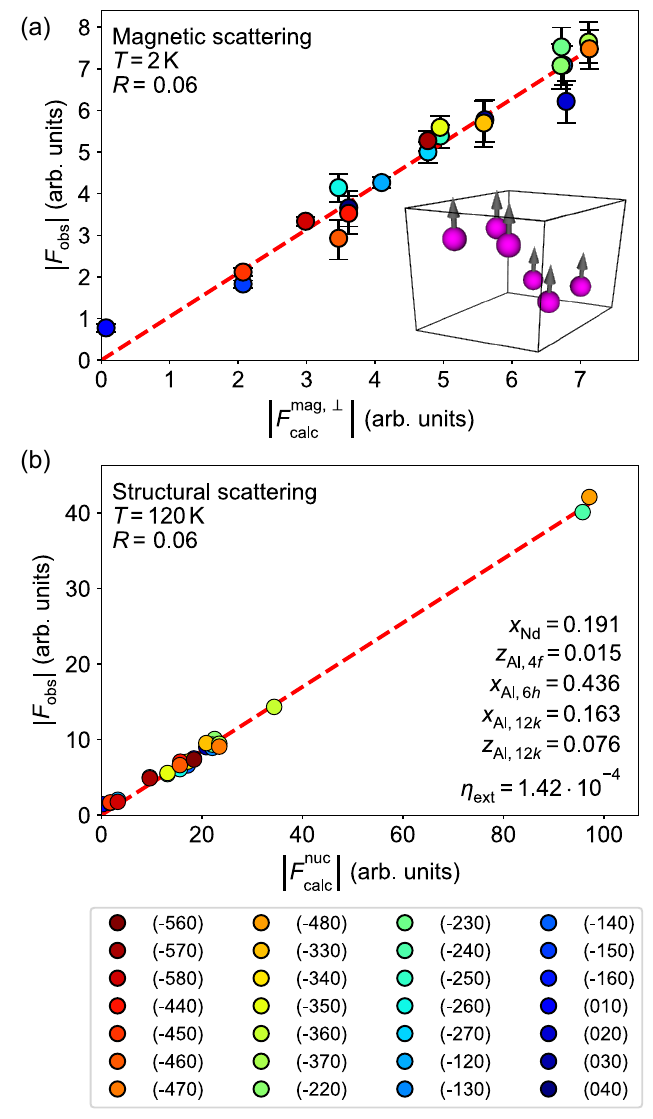}
    \caption[]{
(color online). Refinement of magnetic and crystallographic structure of \NRA{} by comparison of $F_\mathrm{obs}$ and $F_\mathrm{calc}$, the observed and calculated structure factors, respectively. (a) Magnetic structure factor in the $(HK0)$ plane. 
Inset, hex-FM: collinear ferromagnetic structure of magnetic neodymium ions (magenta) in the hexagonal ferromagnetic state. (b) Nuclear scattering related to the crystal structure of \NRA{}. The ratio of the slopes (red lines) of the two plots yields the ordered moment of the Nd ion, $m_c = 2.3\,\mu_\mathrm{B}$. On the right lower side, we give the refined intracell coordinates (in lattice units) for Nd on the $6h$ Wyckoff site and for Al on the $4f$, $6h$, and $12k$ Wyckoff sites. }
    \label{Fig3}
  \end{center}
\end{figure}

\begin{figure}[tb]
  \begin{center}
		\includegraphics[clip, trim=0cm 0cm 0cm 0cm, width=0.7\linewidth]{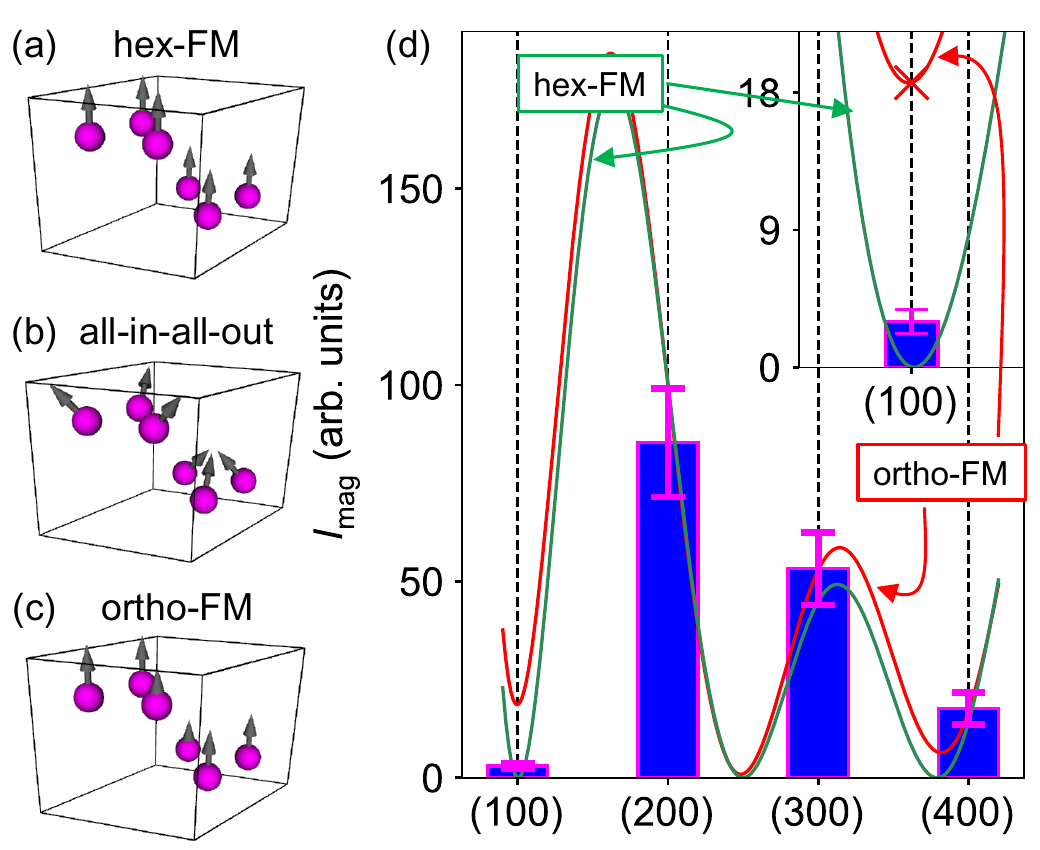}
    \caption[]{
(color online). Magnetic structure models compared along the $(H00)$ line in reciprocal space. (a) Hexagonal ferromagnetic state (hex-FM) with uniform moment length on all sites. Only Nd ions (magenta) are shown. (b) All-in-all-out canted ferromagnetic state, as discussed in the text. (c) Orthorhombic ferromagnetic state (ortho-FM) with two different sites for Nd, as proposed in Ref.~\cite{Gorbunov2016}. (d) Comparison of observed (blue bars with magenta errors) and calculated (red / green lines) magnetic scattering intensity along $(H00)$. For the calculation, we consider the structure factor of a single magnetic unit cell -- thus omitting the Dirac-$\delta$ function part $\delta^{(3)}(\bm{Q}-\bm{G})$, where $\bm{G}$ is a reciprocal lattice vector. For ortho-FM, we assume three equally populated domains. Inset: Magnified view of (d) around the $(100)$ reflection (blue bar), which is about four times weaker than predicted by the model of Ref.~\cite{Gorbunov2016} (cross symbol). Good agreement is obtained with the hex-FM model (green line). The extinction correction ($<4\,\%$) was applied to the experimental intensities.}
    \label{Fig4}
  \end{center}
\end{figure}

\begin{figure*}[tb]
  \begin{center}
		\includegraphics[clip, trim=0cm 0cm 0cm 0cm, width=0.85\linewidth]{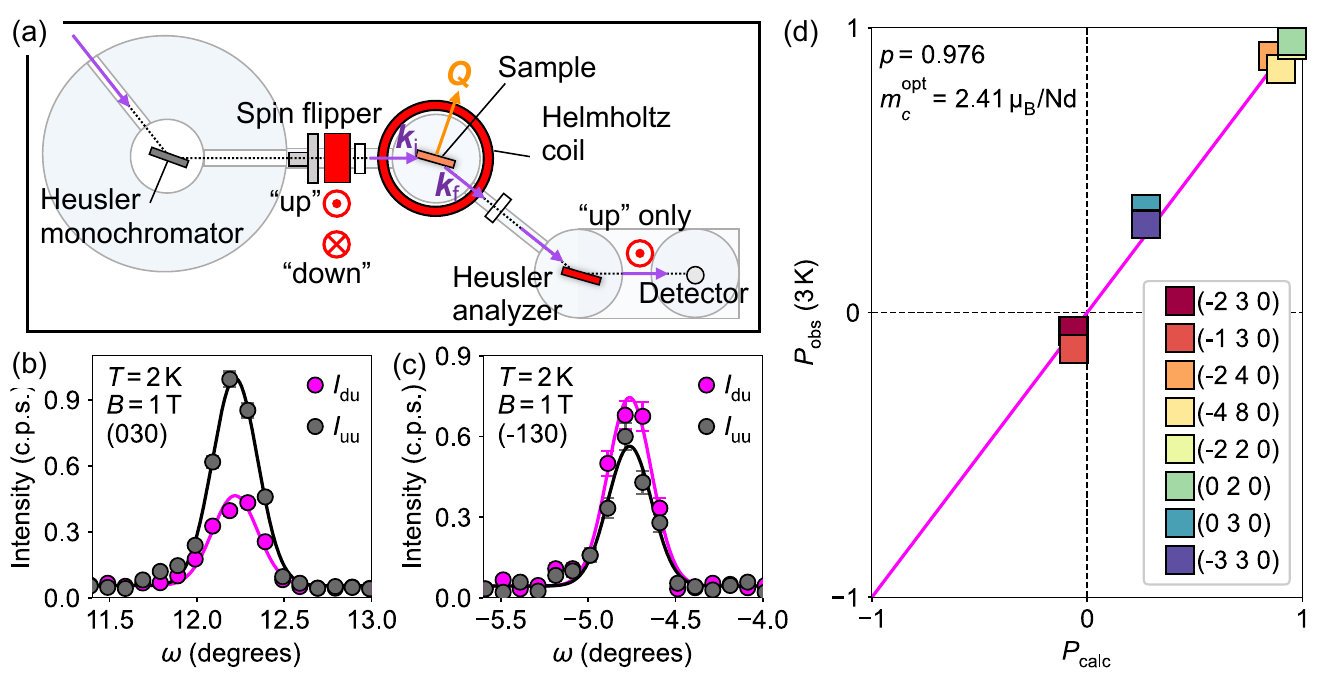}
    \caption[]{
(color online). Polarized neutron scattering on \NRA{} in the $(HK0)$ scattering plane. (a) Experimental geometry. As compared to Fig.~\ref{Fig2}, a spin flipper, Helmholtz coil, and Heusler analyzer for the neutron spin polarization were added (red). The spin of the incoming neutron can be “up” or “down”, whereas the analyzer selects “up” only. (b,c) Rocking scans through two reflections with relatively strong scattering intensity $I_\mathrm{du}$ when the spin flipper is “ON”. (d) Flipping ratio $P_\mathrm{obs} = (I_\mathrm{du}- I_\mathrm{uu})/(I_\mathrm{du}+ I_\mathrm{uu}) $ compared to our calculation result for the hex-FM state, when taking account for neutron beam depolarization $p$. The finite $ I_\mathrm{du}$ for some reflections is attributed to $p$ combined with strong cancellation between magnetic and nuclear scattering in $I_\mathrm{uu}$ (see text).}
    \label{Fig5}
  \end{center}
\end{figure*}

\clearpage
\renewcommand{\thefigure}{A\arabic{figure}}
\setcounter{figure}{0} 

\renewcommand\theequation{A\arabic{equation}}
\setcounter{equation}{0} 

\renewcommand{\thetable}{A\arabic{table}}

\renewcommand\thesection{A \Roman{section}} 
\setcounter{section}{0}

\setcounter{secnumdepth}{3}

\begin{widetext}
\begin{center}
  \textbf{\Large Appendix}
\end{center}
\vspace{1cm}
\end{widetext}

\section{Discussion of spin structure models based on magnetic space groups (mSGs)}
\label{appendix_spacegroups}

\begin{figure}[htb]
\includegraphics[width=0.99\linewidth]{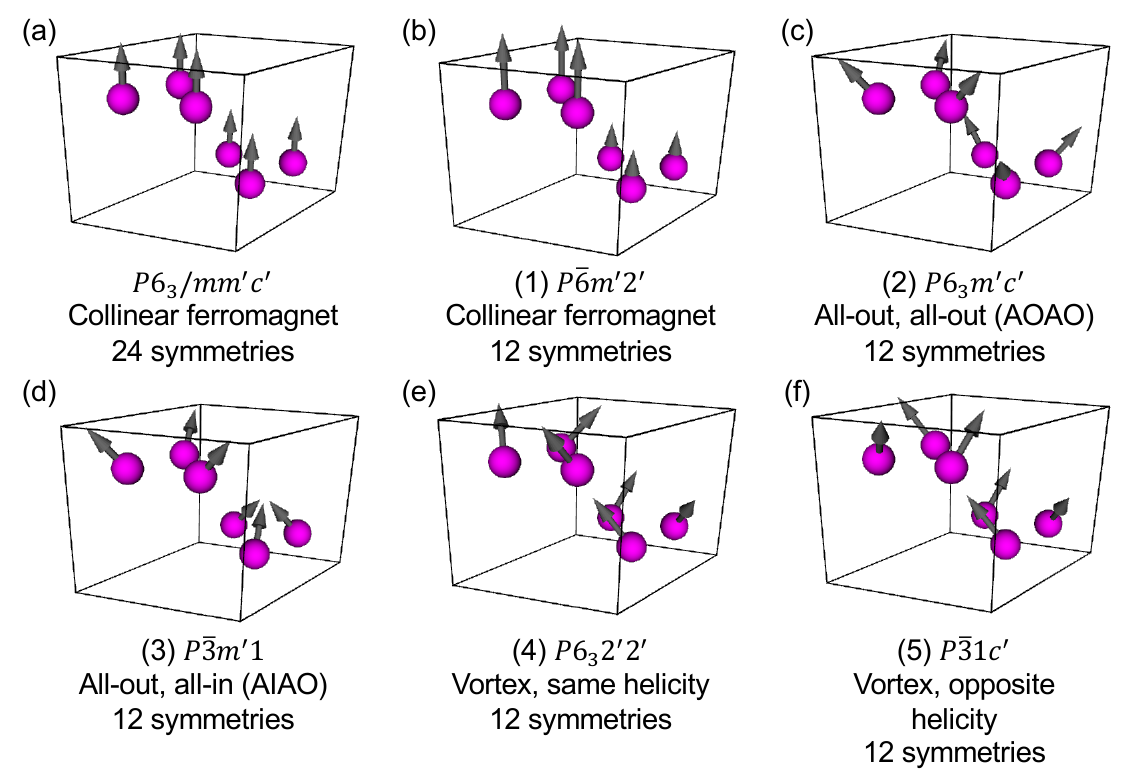}
\caption{Spin structure models for \NRA{} with respective magnetic space group (mSG) symbol and number of unique symmetry operations.}
\label{SFig_spin_models}
\end{figure}

\subsection{Non-magnetic SG of \NRA{}}
We consider the rare earth sublattice of $R_3$Ru$_4$Al$_{12}$ as depicted in Fig.~\ref{Fig1}. The space group (SG) is $P6_3/mmc$, whose full symbol is $P6_3/m\,2/m\,2/c$ with point group $D_{6h}$ or $6/mmm$. The site symmetry of the $R$ ion on the Wyckoff position $6h$ is $mm2$, where a $c$-mirror $\mathcal{M}_c$, a mirror $\mathcal{M}_a$ perpendicular to the $a$-axis, and a $2$-fold rotation axis $C_{2b^*}$ parallel to the $b^*$ direction all pass through the site of the rare earth. Note that the $a$-axis and $b^*$ axis are orthogonal to each other in the hexagonal frame.

\subsection{mSGs for collinear magnetic order in \NRA{}}
\label{appendix_section_mSGs}
Considering the net magnetization of \NRA{} points along the $c$-axis, we first set a magnetic moment, pointing along the $c$-axis, onto site $1$ in Fig.~\ref{Fig1}. Now, $\mathcal{M}_c$ remains intact, $\mathcal{M}_a$ is reduced to $\mathcal{M}_a^\prime$ and $C_{2b^*}$ is reduced to $C_{2b^*}^\prime$. The $2/m$ and $2/c$ symmetry pairs are related by inversion symmetry $\mathcal{I}$.
Thus, the fate of $C_{2a}$ and $\mathcal{M}_{2b^*}$ is tied to whether inversion symmetry is combined with time reversal.
The highest possible space group symmetries in the presence of $m_c$ at site $1$ are $P6_3/m\,2^\prime/m^\prime\,2^\prime/c^\prime$ or $P6_3^\prime/m\,2/m^\prime\,2^\prime/c$. 

The latter mSG has inversion symmetry paired with time reversal, $\mathcal{I}^\prime$. The inversion center of \RRA{} is at $z=1/2$, between the breathing kagome layers, so that $\mathcal{I}^\prime$ results in antiferromagnetic coupling between the $z=1/4$, $z=3/4$ layers and enforces zero net magnetization -- inconsistent with $M \neq 0$ in experiment. The highest remaining mSG with $M\neq0$ along the $c$-axis and moment $m_c$ at site $1$ is $P6_3/m\,2^\prime/m^\prime\,2^\prime/c^\prime$ with $24$ individual symmetry operations. This is a hexagonal, centrosymmetric space group.


\subsection{Further symmetry lowering in \NRA{}, including noncoplanar magnetic orders}

We further lower the symmetry and consider the magnetic subspacegroups of $P6_3/m\,2/m\,2/c$ with $\bm{Q} = 0$ ordering vector and magnetic moment on the $6h$ Wyckoff site. There are six of these:
\begin{enumerate}
\item $P\bar{6}m^\prime 2^\prime$, with $\mathcal{M}_c$ mirror, collinear magnetic order, and $12$ individual symmetries. The moments $m_c$ are identical for all sites in the same layer (e.g. $z=1/4$), whereas, $m_c$ is different on the next layer; c.f. Fig.~\ref{SFig_spin_models}(b). This hexagonal, non-polar and non-chiral space group breaks inversion symmetry $\mathcal{I}$ as well as $\mathcal{I}^\prime$.
\item $P6_3m^\prime c^\prime$, noncoplanar magnetic order with $m_c$ fixed on all six sites in the unit cell and $12$ individual symmetries. This is the "All-out, all-out" (or "All-in, all-in") order in Fig.~\ref{SFig_spin_models}(c) with a hexagonal, polar space group.
\item $P\bar{3}m^\prime 1$, noncoplanar magnetic order with $m_c$ fixed on all six sites in the unit cell and $12$ individual symmetries. This is the "All-in, all-out" order in Fig.~\ref{SFig_spin_models}(d) with a trigonal centrosymmetric space group.
\item $P6_32^\prime 2^\prime$, noncoplanar magnetic order with $m_c$ fixed on all six sites in the unit cell and $12$ individual symmetries. This is the "Vortex, same helicity" order in Fig.~\ref{SFig_spin_models}(e) with a hexagonal, chiral space group.
\item $P\bar{3}1c^\prime$, noncoplanar magnetic order with $m_c$ fixed on all six sites in the unit cell and $12$ individual symmetries. This is the "Vortex, opposite helicity" order in Fig.~\ref{SFig_spin_models}(f) with a trigonal, centrosymmetric space group.
\item $Cm^\prime c^\prime m$, orthorhombic with $8$ individual symmetries. This is the mSG proposed in Ref.~\cite{Gorbunov2016}. 
\end{enumerate}


\section{Flipping ratio in polarized neutron scattering}
\label{appendix:polarized_neutron}
The half-$\lambda$ correction is applied to the polarized neutron scattering data based on the assumption that the higher harmonic is fully spin-depolarized. Calculating the flipping ratio 
\begin{equation}
P_\mathrm{obs} = \frac{I_\mathrm{du}^\mathrm{obs} - I_\mathrm{uu}^\mathrm{obs}}{I_\mathrm{du}^\mathrm{obs} + I_\mathrm{uu}^\mathrm{obs}}
\end{equation}
and requiring that this value should be the same for all reflections at $T = 120\,$K, we find that only the $(110)$ represents a significant outlier. Therefore, we subtract $I_\mathrm{uu} \rightarrow I_\mathrm{uu}-I_{\lambda/2}$ and $I_\mathrm{du} \rightarrow I_\mathrm{du}-I_{\lambda/2}$ with $I_{\lambda/2}$ chosen so as to make $P$ uniform at high temperature. We also assume that $I_{\lambda/2}$ is unchanged upon entering the ferromagnetic state.

Next, to quantitatively define the polarization $p$ of the neutron spins at the sample, we introduce the structure factors resolved by incoming and outgoing neutron spin state (in the present geometry, the scattering plane is perpendicular to the $c$-axis of the crystal),
\begin{widetext}
\begin{align}
F_\mathrm{uu}^\mathrm{calc} &= 
\sum_{j\in \mathrm{u.c.}} b_{\mathrm{Ru/Al}} 
   \exp\!\left(\imath \bm{Q}\cdot \bm{r}_j \right)
+ \sum_{j\in \mathrm{u.c.}} 
   \Big[ b_{\mathrm{Nd}} - 2.7\, f_{\mathrm{mag}}(Q)\, m_j^c \Big] 
   \exp\left(\imath \bm{Q}\cdot \bm{r}_j \right), \\
F_\mathrm{dd}^\mathrm{calc} &= 
\sum_{j\in \mathrm{u.c.}} b_{\mathrm{Ru/Al}} 
   \exp\left(\imath \bm{Q}\cdot \bm{r}_j \right)
+ \sum_{j\in \mathrm{u.c.}} 
   \Big[ b_{\mathrm{Nd}} + 2.7\, f_{\mathrm{mag}}(Q)\, m_j^c \Big] 
   \exp\left(\imath \bm{Q}\cdot \bm{r}_j \right)\\
\bm{F}_{\rm ud}^\mathrm{calc} &= \bm{F}_{\rm du}^\mathrm{calc} = -2.7 \cdot  \sum_{j} \bm{m}_j^{ab}\,f_{\mathrm{mag},j}(Q)\exp(\imath {\bm Q}\cdot {\bm r}_j)\label{eq:polarized_FSF}
\end{align}
where $Q = \left|\bm{Q}\right|$ and $\bm{m}^{ab}_j$ is the component of the magnetic moment at site $j$ in the $ab$ basal plane of hexagonal \NRA{} (now, not normalized to unity). The calculated scattering intensities for "spin flip" and "non-spin flip" events for reflections in the $(HK0)$ scattering plane are
\begin{align}
I_{\rm du}^{\rm calc}/\Phi&=p^2 \cdot {|\bm{F}_{\rm du}^{\rm calc}|}^2 + (1-p)^2 \cdot {|\bm{F}_{\rm ud}^{\rm calc}|}^2 
	+ (1-p)p \cdot {|F_{\rm dd}^{\rm calc}|}^2 + p(1-p) 	\cdot {|F_{\rm uu}^{\rm calc}|^2}\label{eq:Idu}\\
I_{\rm uu}^{\rm calc}/\Phi&=p^2 \cdot {|F_{\rm uu}^\mathrm{calc}|}^2 + (1-p)^2 \cdot {|F_{\rm dd}^\mathrm{calc}|}^2 
	+ (1-p)p \cdot {|\bm{F}_{\rm du}^{\mathrm{calc},\perp}|}^2 + p(1-p) \cdot {|\bm{F}_{\rm ud}^{\mathrm{calc},\perp}|^2}\label{eq:Iuu}.
\end{align}
\end{widetext}
For reflections with strong nuclear intensity, such as $(-240)$ and $(-480)$, $\bm{F}_\mathrm{ud}^\mathrm{calc}$ in Eq.~\eqref{eq:polarized_FSF} is weak and we can calculate $p$ by comparing $P_\mathrm{calc} $ and $P_\mathrm{obs}$. Experimentally, we find $p = 0.976$ at both $T=2\,$K, $120\,$K from $(480)$ with large $Q$ and consequently $f^\mathrm{mag}\rightarrow 0$; this value of $p$ gives a consistent description of our data. 

We note that Eq.~\eqref{eq:Idu} produces finite "spin flip" intensity $I_\mathrm{du}$ even for a fully $c$-collinear magnetic order, as experimentally observed in Fig.~\ref{Fig5}. In fact, $I_\mathrm{du}$ can be comparable to $I_\mathrm{uu}$ due to partial cancellation between nuclear and magnetic scattering in Eq.~(\ref{eq:Iuu}). Comparable intensity of spin-flip and non-spin flip scattering is observed at $(-130)$, for example.


\begin{table*}[tb]
\centering
\caption{Refined structural parameters of \RRA{} ($R=\,$Nd, Gd) from this work and from previous reports. The experimental neutron diffraction results for Sample~2 and Sample~3 are discussed in the Supplementary Material.}
\label{tab:refined_parameters}
\setlength{\tabcolsep}{7pt}
\renewcommand{\arraystretch}{1.3} 
\begin{tabular}{|l | c | c | c | c | c | c | c|}
\hline
Material & Method & Sample & $x_\mathrm{Nd}$ & $z_{\mathrm{Al},4f}$ & $x_{\mathrm{Al},6h}$ & 
   $x_{\mathrm{Al},12k}$ & $z_{\mathrm{Al},12k}$ \\
\hline
\NRA{} & neutron & Sample~1 & 0.191 & 0.015 & 0.436 & 0.163 & 0.076 \\
\NRA{} & neutron & Sample~2 & 0.191 & 0.012 & 0.438 & 0.162 & 0.077\\
\NRA{} & neutron & Ref.~\cite{Gorbunov2016} & 0.19165(3) & 0.00866(3) & 0.43841(1) & 0.16231(1) & 0.0754(2) \\
\GRA{} & in-house X-ray & Ref.~\cite{Hirschberger2019} & 0.19287 & 0.01550 & 0.44155 & 0.16350 & 0.07657 \\
\hline
\end{tabular}
\end{table*}

\begin{table*}
    \caption{\textbf{\NRA{} single crystals used in this work.} We provide approximate outer dimension of single crystal pieces and their precise weight. Sample~2 includes an outer layer of amorphous material (volume fraction $<5\,\%$) produced in the crystal growth. This outer layer was removed by polishing for Sample~1 and Sample~2. The experimental neutron diffraction results for Sample~2 and Sample~3 are discussed in the Supplementary Material.
    }
    \begin{tabular}{|c|c|c|c|c|c|c|c|c|}\hline
         Sample& dimensions  & weight & experiment & neutron energy $E_i$  \\  
         &(mm) & (mg) & (beamline) & (meV) \\ \hline \hline 
         
        Sample 1 & $1.6 \times 3.3 \times 3.4$ & $104.13$ & JRR-3, PONTA-5G &  $34.05$\\ 
        \hline
        Sample 2 & $0.88 \times 3.4 \times 3.4$ & $23.67$ & JRR-3, PONTA-5G & $34.05$\\
        \hline
        Sample 3 & $2.0 \times 3.9  \times 5.7$ 
        & $450$ &  ORNL, HB-1 & $13.5$\\ \hline 
          
    \end{tabular}

    \label{Table_Samples}
\end{table*}
\color{black}

\end{document}